\documentclass[a4paper,fleqn,usenatbib,useAMS,usefloat]{mnras}
\usepackage{mathptmx}
\usepackage[normalem]{ulem}
\usepackage[T1]{fontenc}
\usepackage{ae,aecompl}
\usepackage{longtable}

\usepackage{graphicx}	
\usepackage{epstopdf}
\usepackage{xcolor}
\usepackage{amsmath}	
\usepackage{amssymb}	
\usepackage{subfig}
\usepackage{ulem}
\usepackage{booktabs}
\usepackage{float}
\usepackage{array}
\usepackage{caption}
\usepackage{scalerel}
\title[Corrected amplitudes of ellipsoidal modulation]
{Search for Dormant Black Holes in Ellipsoidal Variables I. Revisiting the Expected Amplitudes of the Photometric Modulation}
\author[Gomel et al.]{
\newauthor{Roy Gomel, Simchon Faigler and Tsevi Mazeh} 
\\
{School of Physics and Astronomy, Faculty of Exact Sciences,}\\
{Tel Aviv University, Tel Aviv  69978, Israel}
}
\date{Accepted XXX. Received YYY; in original form ZZZ}

\pubyear{2019}

\begin{document}
\label{firstpage}
\pagerange{\pageref{firstpage}--\pageref{lastpage}}
\maketitle

\begin{abstract}
Ellipsoidal variables present light-curve modulations caused by stellar distortion, induced by tidal interaction with their companions. 
An analytical approximated model of the ellipsoidal modulation is given as a discrete Fourier series by Morris and Naftilan 1993 (MN93). 
Based on numerical simulations using the PHOEBE code we present here updated amplitudes of the first three harmonics of the model. The expected amplitudes are given as a function of the mass ratio and inclination of the binary system and the fillout factor of the primary---the ratio between the stellar radius and that of its Roche lobe. The corrections can get up to $30$\% relative to the MN93 model for fillout factors close to unity.
The updated model can be instrumental in searching for short-period binaries with compact-object secondaries in large  data sets of photometric light curves. As shown in one OGLE light-curve example, the minimum mass ratio can be obtained by using only the amplitudes of the three harmonics and an estimation of the stellar temperature. High enough amplitudes can help to identify binaries with mass ratios larger than unity, some of which might have compact companions.
\end{abstract}

%
\section{Introduction}

In recent years a few ground-based and space-mission surveys, such as OGLE \citep{udalski15}, HATNet \citep{hartman04}, HATSouth \citep{bakos13}, WASP \citep{pollacco06}, {Kepler} \citep{borucki10, koch10}, Gaia \citep{Gaia2016}, TESS \citep{ricker15}, Pan-STARRS \citep{kaiser10} and Catalina \citep{drake14}, have obtained millions of stellar light curves with many data points of high precision. 
These data give the opportunity to discover a large number of non-eclipsing close binaries \citep[e.g.,][]{soszy04, faigler12, soszy16, shporer17},
 based on their ellipsoidal modulation,
 caused by stellar distortion induced by tidal interaction with their companions.

In most of the close systems, 
the unseen companions are faint main-sequence stars with small masses  
\citep[e.g.,][]{raghavan10, moe17}. 
However, some binaries have unseen compact companions which are more massive than the observed components, 
black-hole (BH) companions in particular.
For example, some 20 dynamically confirmed stellar BHs are known to reside in close binary systems with low-mass stellar companions. These are, along with Cyg X-1 and several other high-mass binary candidates, the only confirmed stellar-mass BHs in the Galaxy
\citep[see, for example,][]{corral16}.
All stellar-mass BHs known so far have been discovered by their X-ray emission, 
due to either mass transfer from a low-mass 
(mostly K--F star) companion overflowing its Roche lobe (BH-LMXBs), 
or accretion from a stellar wind coming from a high-mass  (O--B star) companion (BH-HMXBs) 
\citep[e.g.,][]{fabian89, remillard06, orosz07, ziolkowski14}.
According to the commonly accepted model, the BH-LMXB outbursts are due to some disk instabilities 
\citep[][]{lasota01} 
that modulate the accretion rates onto the BH. Between eruptions, these systems are barely detectable, because a substantial part of the energy generated by the small mass flow is not radiated but stored as thermal energy in their discs.  
Thus, many BH-LMXB remained undetected, 
because they have been in their quiescent state when observed by the X-ray surveys 
%
%
\citep[e.g.,][]{ritter02, cackett05, knevitt14}.

A much larger fraction of BHs with low-mass stellar companions are not detected yet
because their optical counterparts are well within their Roche lobes, so mass is not transferred and X-rays are not generated, making these systems dormant BHs 
\citep[see discussion on the frequency of such systems by][]{breivik17, mashian17, yamaguchi18, shao19, yi19, wiktorowicz19, shikauchi20}.
We are aiming to discover some of the short-period dormant systems by their ellipsoidal modulations.

To identify the ellipsoidal variables and distinguish between them and other stellar variables \citep[e.g.,][]{pojmanski02,soszy11a,soszy11b,soszy13}, and to identify the systems with massive companions in particular,  one needs a reliable approximation of the expected ellipsoidal modulation that can be applied to large data sets and yield sound estimate of the companion mass. 
\citet[][MN93]{morris93} 
classical work, based on \citet{kopal59} approach
\citep[see also][]{russell45},
derived an analytical approximation of the ellipsoidal modulation for {\it circular} orbits 
\citep[see a very recent extension by][to eccentric orbits]{engel20}. 
Their work was based on an expansion of the tidal interaction between the primary and the secondary in a power series of $R_1/a$, where $R_1$ was the mean radius of the primary and $a$ was the semi-major axis of the system. 

The resulting model of the ellipsoidal modulation was presented by MN93 in harmonics of the orbital phase, each of which was given by  a power series of $R_1/a$, up to the fifth power. MN93 gave closed formulae for the amplitudes of the first four
harmonics as a function of the mass ratio, orbital inclination and $R_1/a$ of the binary. 
Indeed, MN93 was used successfully in many studies  \citep[e.g.,][]{zucker07,mazeh10,faigler15,sullivan15,parsons17, masuda19}.
MN93 approximation works well for cases for which
$R_1/a$ has a small value, but becomes inaccurate for cases with $R_1$ close to the Roche-lobe radius. In such cases, the oval shapes of the stars 
have to be calculated numerically by equipotential surfaces inside, but close, to the Roche-surface limit.

With the advance of computational power, a few numerical codes have been developed \citep[e.g.,][]{hill70,wilson71,bochkarev79,orosz00} to simulate the ellipsoidal modulation of close binary systems. The basic idea is to derive numerically the stellar equipotential surface,  divide this surface to small discrete elements, derive the luminosity of each element, and obtain the total luminosity by summing up the light coming from the whole stellar surface. These simulated light curves yield a better approximation of the ellipsoidal modulation 
\citep[e.g.,][]{mcclintock86, casares93, shahbaz93}, provided the assumptions behind the codes about stellar structure and atmosphere are accurate enough.
However, the calculation of each light curve requires considerable CPU time,  and therefore these codes are inapplicable for analyzing millions of light curves with unknown periods and orbital elements, even with the growing speed of nowadays computers.

As a first step of the search for dormant BHs in close binaries, we present here easy-to-use pre-calculated  amplitudes of the first three harmonics of the ellipsoidal modulation as a function of
the stellar mass, radius and effective temperature of the primary, the orbital period and inclination, and the mass ratio of the binary.
This was done by
applying the PHOEBE 2.1 (PHysics Of Eclipsing BinariEs) software package 
\citep{prsa05,prsa16,horvat18,jones19}, 
%
that was 
successfully used in many studies \citep[e.g.,][]{torres10, prsa11, eastman13, jones17, shporer17}. 
We applied PHOEBE to systems chosen from a dense grid of the parameter space, presenting our results as a correction factor to the MN93 approximation for the amplitudes of the first three harmonics of the modulation.


A key parameter that determines  the ellipsoidal modulation is the fillout factor of the primary --- the primary volume-averaged radius divided by the Roche-lobe volume-averaged radius
\citep{kopal59,paczynski71,eggleton83}. Therefore, the corrections  are presented here for a grid of values of the fillout factor, the mass ratio and the orbital inclination. 
Finally, based on our grid points we have found approximate simple expressions for the correction factor of the amplitudes of the second and third harmonics for any given system.   
For the first harmonic, we present a Python code to obtain the corrected amplitudes based on a linear interpolation between the grid points for different main-sequence primaries and optical bands. 

Section \ref{sec:MN93} presents the MN93 approximation and its limitations, Section \ref{sec:analytic} derives the correction terms relative to MN93 expressions, Section \ref{sec:test} considers one example, demonstrating how our analysis can work for one specific OGLE system, and Section \ref{sec:summary} summarizes our results.

\section{The Morris and Naftilan approximation}
\label{sec:MN93}

Let us consider a binary system of two stars, with masses $M_1$ and $M_2$, for which we observe the light coming from the primary $M_1$ star only. We are interested in the relative ellipsoidal modulation in some known optical band, caused by the tidal interaction with the secondary.

The approximation for the primary-star ellipsoidal modulations of the first three harmonics is given by MN93, assuming 
tidally-locked ellipsoidal variables in circular orbits and $R_1/a \ll 1$, and adopting linear limb- and gravity-darkening laws. Their equation is

%
\begin{equation}
\begin{split}
&\frac{\Delta L}{\overline{L}} = \
\frac{\alpha_\mathrm{1}}{\overline{L}/L_0}
\left(\frac{R_1}{a}\right)^4 
q
\left( 4 \sin i - 5 \sin^3 i \right) \ \cos \phi \ \\
&-\frac{1}{\overline{L}/L_0}\left(\alpha_\mathrm{2}
\left(\frac{R_1}{a}\right)^3 
q
\sin^2 i
+\beta_\mathrm{2}
\left(\frac{R_1}{a}\right)^5 
q
\left(6\sin^2 i - 7\sin^4 i\right) \right) \
\cos 2\phi \ \\
&-\frac{5}{3}
\frac{\alpha_\mathrm{1}}{\overline{L}/L_0}
\left(\frac{R_1}{a}\right)^4 
q
\sin^3 i \ \cos 3\phi \ ,
\end{split}
\label{eq:ellip1}
\end{equation}
where $\phi$ is the orbital angle, with $\phi$=0 defined to be 
at superior conjunction. Here the binary mass ratio is
$q=M_2/M_1$,
the ellipsoidal coefficients are defined by
\begin{equation}
\alpha_\mathrm{1}=\frac{15u(2+\tau)}{32(3-u)} \ , \\
\alpha_\mathrm{2}=\frac{3(15+u)(1+\tau)}{20(3-u)} \ , \\ 
\beta_\mathrm{2}=\frac{15(1-u)(3+\tau)}{64(3-u)} \ , \\
\label{eq:alpha}
\end{equation}
and the average luminosity of the star is given by
\begin{equation}
\overline{L}=L_0\{1 + \frac{1}{9} \alpha_\mathrm{2} \left(\frac{R_1}{a}\right)^3(2+5q)\left(2-3sin^2 i\right)\} \ , \\ 
\label{eq:Lav}
\end{equation}
with $L_0$ being the stellar brightness with no secondary at all. \\
In the above equations 
$R_1$ is the volume-averaged radius of the primary, $a$ is the binary semi-major axis, $i$ is the orbital inclination, and $u$ and $\tau$ are the linear limb- and gravity-darkening coefficients of the primary.

Using the linear limb- and gravity-darkening coefficients of \cite{claret11} for stars with an effective temperature between $4000$--$7000$ K, Sun-like gravity and zero metallicity, we find that the $\alpha_2$ coefficient is typically between $1$--$2$ and $\alpha_1$, $\beta_2$ are in the range $0$--$0.4$.

Under this simple model of circular, synchronous and aligned components, the distorted surface of the primary star is symmetrical with respect to orbital angles $\phi=0$ and $\pi$ for any inclination. Thus, the light curve of the ellipsoidal has to be symmetric around these angles and contain only cosine terms in its Fourier expansion.

As can be seen in the equations above, the leading term in the expansion is $(R_1/a)^3$, which appears in the expression of the second-harmonic amplitude only. 
The amplitude is therefore approximately proportional to $R_1^3$, everything else being equal. The dependence of $q$ is more subtle. All three amplitudes  depend linearly on $q$, if $a$ is known. However, if one uses the orbital period and the primary mass of the binary as the known parameters, which is often the case, then the $q$ dependence is hidden, because for a given period and primary mass $a^3\propto 1+q$. 

The expressions for the amplitudes of the first and third harmonics have $R_1/a$ to the fourth power, one factor higher than in the second harmonic expression. In most cases $R_1/a$ is much smaller than 1. 
Furthermore, the $\alpha_2$ coefficient is typically larger than $\alpha_1$.
Thus, the second-harmonic amplitude is typically an order of magnitude larger than the other two terms of Equation~(\ref{eq:ellip1}),
which gives the characteristic double-peaked appearance to the light curve. 

Note that Equation~(\ref{eq:Lav}) indicates that $\overline{L} \neq L_0$ even when the secondary's mass tends to zero. This is explained by the factor $2+5q$ which is, in fact, a sum of two effects. 
The first effect, which is proportional to  $2(1+q)$,  is the outcome of the centrifugal force that distorts the stellar shape. This effect comes from the fact that the primary is at rest in the rotating frame and therefore at work even for $q\to 0$. 
The second effect, which is proportional to  $3q$,  is due to distortion of the stellar shape by the secondary-star tidal forces, thus vanishing at $q\to 0$.

\subsection{The dependence on the fillout factor}

Equation~(\ref{eq:ellip1}) can be expressed by the Roche-lobe radius as
%
\begin{equation}
\begin{split}
&\frac{\Delta L}{\overline{L}} = \
\frac{\alpha_\mathrm{1}}{\overline{L}/L_0}
\left(\frac{R_{\rm Roche,1}}{a}\right)^4
f^4 
q
\left( 4 \sin i - 5 \sin^3 i \right) \ \cos \phi \ \\
&-\frac{1}{\overline{L}/L_0} \biggl( \biggr. \alpha_\mathrm{2}
\left(\frac{R_{\rm Roche,1}}{a}\right)^3
f^3
q
\sin^2 i \ \\
&+\beta_\mathrm{2}
\left(\frac{R_{\rm Roche,1}}{a}\right)^5 
f^5
q
\left(6\sin^2 i - 7\sin^4 i\right) \biggl. \biggr) \
\cos 2\phi \ \\
&-\frac{5}{3}
\frac{\alpha_\mathrm{1}}{\overline{L}/L_0}
\left(\frac{R_{\rm Roche,1}}{a}\right)^4 
f^4
q
\sin^3 i \ \cos 3\phi \ ,
\end{split}
\label{eq:ellipr1}
\end{equation}
%
with the Roche-lobe fillout factor and the average luminosity defined by
\begin{equation}
\begin{split}
&f=\frac{R_1}{R_{\rm Roche,1}} \ , \ \\
&\overline{L}=L_0\{1 + \frac{1}{9} \alpha_\mathrm{2} \left(\frac{R_{\rm Roche,1}}{a}\right)^3f^3(2+5q)\left(2-3sin^2 i\right)\} \ . \\ 
\end{split}
\end{equation}

Using the \cite{eggleton83} approximation for the volume-averaged Roche-lobe radius, which is accurate to $1\%$ over the entire $q$ range,
%
\begin{equation}
E(q) \equiv \frac{0.49 q^{-2/3}  }{0.6 q^{-2/3} + \ln(1+q^{-1/3}) } \ \approx \ \frac{R_{\rm Roche,1}}{a} \ , 
\label{eq:egg}
\end{equation}
we get
\begin{equation}
\begin{split}
&\frac{\Delta L}{\overline{L}} = \
\frac{\alpha_\mathrm{1}}{\overline{L}/L_0}
E^4(q)
f^4 
q
\left( 4 \sin i - 5 \sin^3 i \right) \ \cos \phi \ \\
&-\frac{1}{\overline{L}/L_0}\left(\alpha_\mathrm{2}
E^3(q)
f^3
q
\sin^2 i
+\beta_\mathrm{2}
E^5(q)
f^5
q
\left(6\sin^2 i - 7\sin^4 i\right) \right) \
\cos 2\phi \ \\
&-\frac{5}{3}
\frac{\alpha_\mathrm{1}}{\overline{L}/L_0}
E^4(q)
f^4
q
\sin^3 i \ \cos 3\phi \ ,
\end{split}
\label{eq:ellipegg}
\end{equation}
%
with
\begin{equation}
\overline{L}=L_0\{1 + \frac{1}{9} \alpha_\mathrm{2} E^3(q)f^3(2+5q)\left(2-3sin^2 i\right)\} \ . \\ 
\label{eq:Lavegg}
\end{equation}
%
The (semi-)amplitudes of the $\{\cos n \phi;\, n = 1,2,3\}$  harmonics 
in Equation~(\ref{eq:ellipegg}) will be denoted by $A_{\rm n,\scaleto{MN}{5pt}}$. 

 In the new expressions, the dependence on the stellar radius is through the fillout factor, as $f\propto R_1$. The dependence on $q$ is through the Eggleton expression $E(q)$, which is less apparent.

As pointed out above, MN93 analysis assumes $R_1/a$ is small and therefore $f$ has a small value. For $f$ near unity the MN93 approximation can significantly deviate, up to $\sim 50$\%,
  from the actual ellipsoidal variability  \citep[e.g.,][]{bochkarev79}. It is therefore important to numerically derive the correct value of the harmonic amplitudes, as done in the next section.

\begin{figure*}
	\centering
	\resizebox{16cm}{6cm}
	{
		\includegraphics{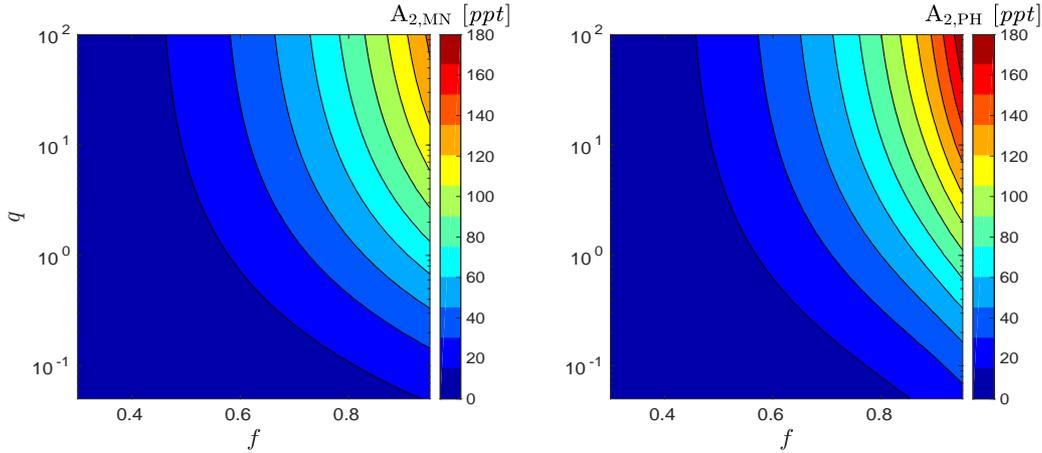}  
	}
	\caption{{\it Left:} Expected  $V$ amplitude of the second harmonic according to MN93 approximation, as a function of the Roche-lobe fillout factor of the primary $f$ and the binary mass-ratio $q$. This was derived for a binary with Sun-like star and $\sin i=1$, with no contribution from the secondary and no beaming effect. Contour lines with equally-spaced amplitudes are drawn with solid lines from $15$ ppts upwards in steps of $15$ ppts. 
		{\it Right:} Expected second-harmonic amplitude derived by PHOEBE.  Its maximum can reach up to $\sim 180$ ppt, $\sim 30\%$ larger than the approximated values. 
	}
	\label{fig:mn93}
\end{figure*}

\section{ Correcting the MN93 amplitude }
\label{sec:analytic}

\subsection{Comparing the PHOEBE models with the MN93 approximation}

To better estimate the ellipsoidal effect we used  the PHOEBE\footnote{http://phoebe-project.org} (PHysics Of Eclipsing BinariEs) software package \citep{prsa05,prsa16,horvat18,jones19} to simulate light curves of close binaries with the ellipsoidal modulation only, assuming  a circular orbit, with an aligned and synchronous rotation of the primary. 

The simulations were run for a Sun-like primary (see below a discussion for an extension  of the model to different stars) and a compact-object secondary, so we set $T_{\rm eff}=0$ and $R=10^{-4}R_{\odot}$ for the secondary.
In addition, we disabled the beaming (sometimes called Doppler boosting) effect of the two components in the simulated light curve, as this paper focuses on the ellipsoidal effect only.  The light curves were simulated in the $V$ band using the \cite{castelli04} atmospheric models for the primary star. In the simulation, the primary star was divided into $2\cdot10^5$ surface elements.

The bolometric gravity-darkening exponent $\beta_1$ used by PHOEBE was taken from \cite{claret04} and the linear limb-darkening coefficient from \cite{claret11} for the same stellar parameters. PHOEBE light curves simulated with different limb-darkening laws for the primary presented a scatter of a few percent in the Fourier coefficients of the first three harmonics, but a change in the gravity-darkening exponent value may introduce more significant variations. 

After normalizing the PHOEBE light curve to an average value of $1$, we fitted the modulation with three harmonics,
with $\phi=0$ at superior conjunction,
 resulting in three Fourier coefficients, $\{A_{\rm n,\scaleto{PH}{5pt}};\, n = 1,2,3\}$.

To compare the PHOEBE results with the MN93 formulae, we used Equation~(\ref{eq:ellipegg}) to obtain  the first three harmonics, $\{A_{\rm n,\scaleto{MN}{5pt}};\, n = 1,2,3\}$ for the same binary parameters,
using the linear limb- and gravity- darkening coefficients of a Sun-like star with $T_{\rm eff}$ of $5780$\,K from \cite{claret11}. Here again, the result is more sensitive to the value of the gravity-darkening coefficient (see $\alpha_2$ in Eq.~\ref{eq:alpha}). 

One possible comparison is presented in Fig.~\ref{fig:mn93}, where we plot the expected second-harmonic amplitudes of the MN93 (left panel) and PHOEBE (right panel) models side by side. This is done by 2D surfaces as a function of the Roche-lobe fillout factor $f$ and the binary mass-ratio $q$
for $\sin i=1$.

One can see in both panels that the amplitude is  of the order of $50$ parts-per-thousand (ppt), or $\sim5$\%, for  a  binary of $q\sim1$ and $f\sim0.75$. The amplitude is 
rising monotonically with $f$ and $q$, and can reach, in extreme cases, up to $\sim 140$ ppt in the MN93 approximation and $\sim 180$ ppt in the PHOEBE model.
This indicates that the difference between the two models can reach, for high $q$ and $f$, up to $\sim30$\%.
As we are interested in this part of the parameter space, an easy-to-use correction factor of the MN93 approximation can be of much use.

The essential role of such correction factor is emphasized by the amplitude derived for the second harmonic of OGLE-BLG-ELL-007730 of Fig.~\ref{fig:ogle}, 
$A_{\rm 2,obs}\simeq 0.14$, which clearly put OGLE-BLG-ELL-007730 in the upper-right corner of the amplitude plots. This example will be discussed further later.


\subsection{Presenting the correction with simple expressions}

\begin{figure*} 
	\centering
	\resizebox{17cm}{17.0cm}
	{		\includegraphics{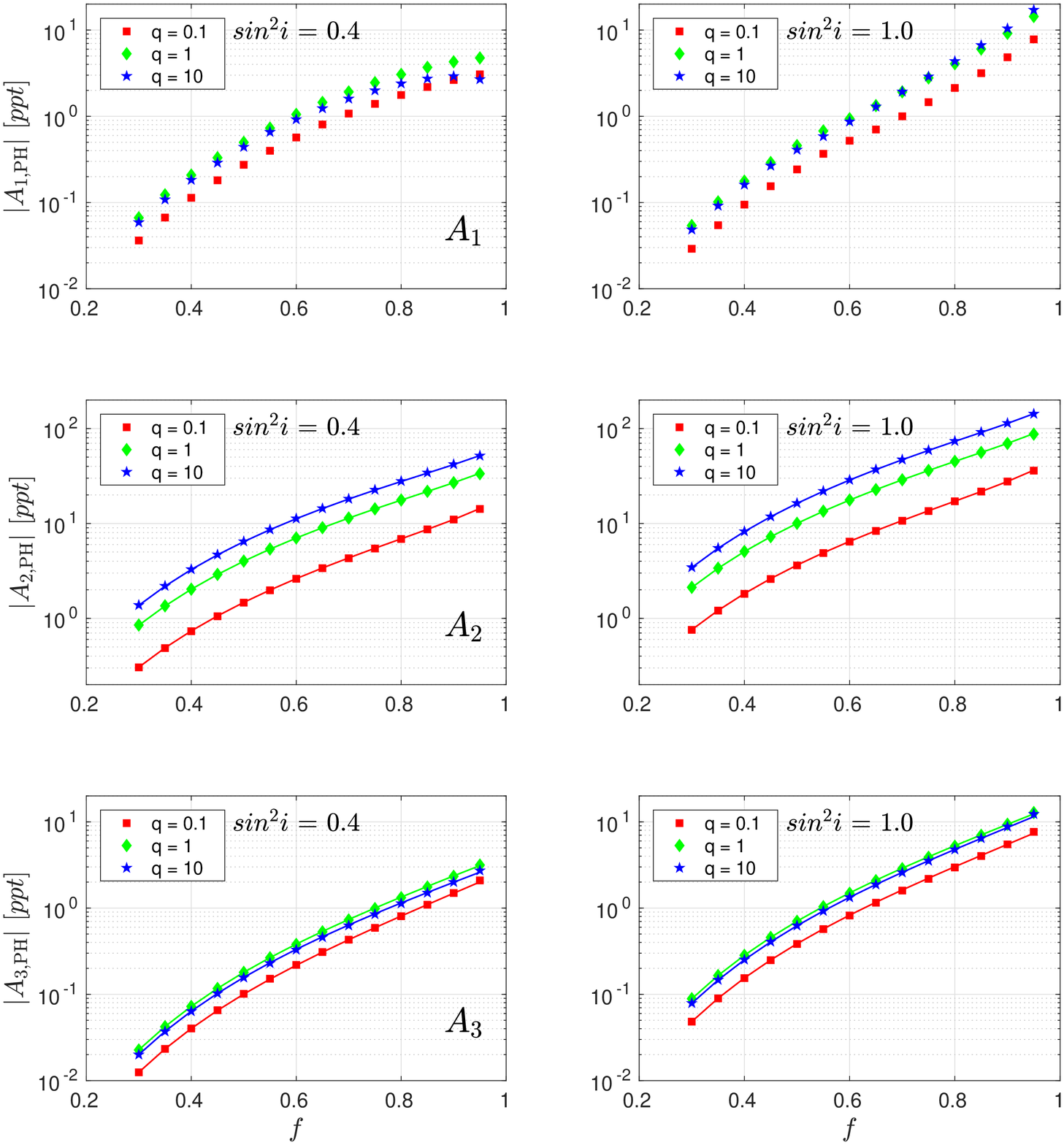}  	}
	\caption{Amplitudes of the PHOEBE model as a function of fillout factor for a Sun-like star. Amplitudes are given for two inclinations and three mass-ratio values. Top panel shows the amplitudes for the first harmonic. The two lower panels present the second- and third-harmonic amplitudes (points) and their approximations (solid line).
	}
	\label{fig:acPHOEBE}
\end{figure*}

\begin{figure*} 
	\centering
	\resizebox{17cm}{17.0cm}
	{\includegraphics{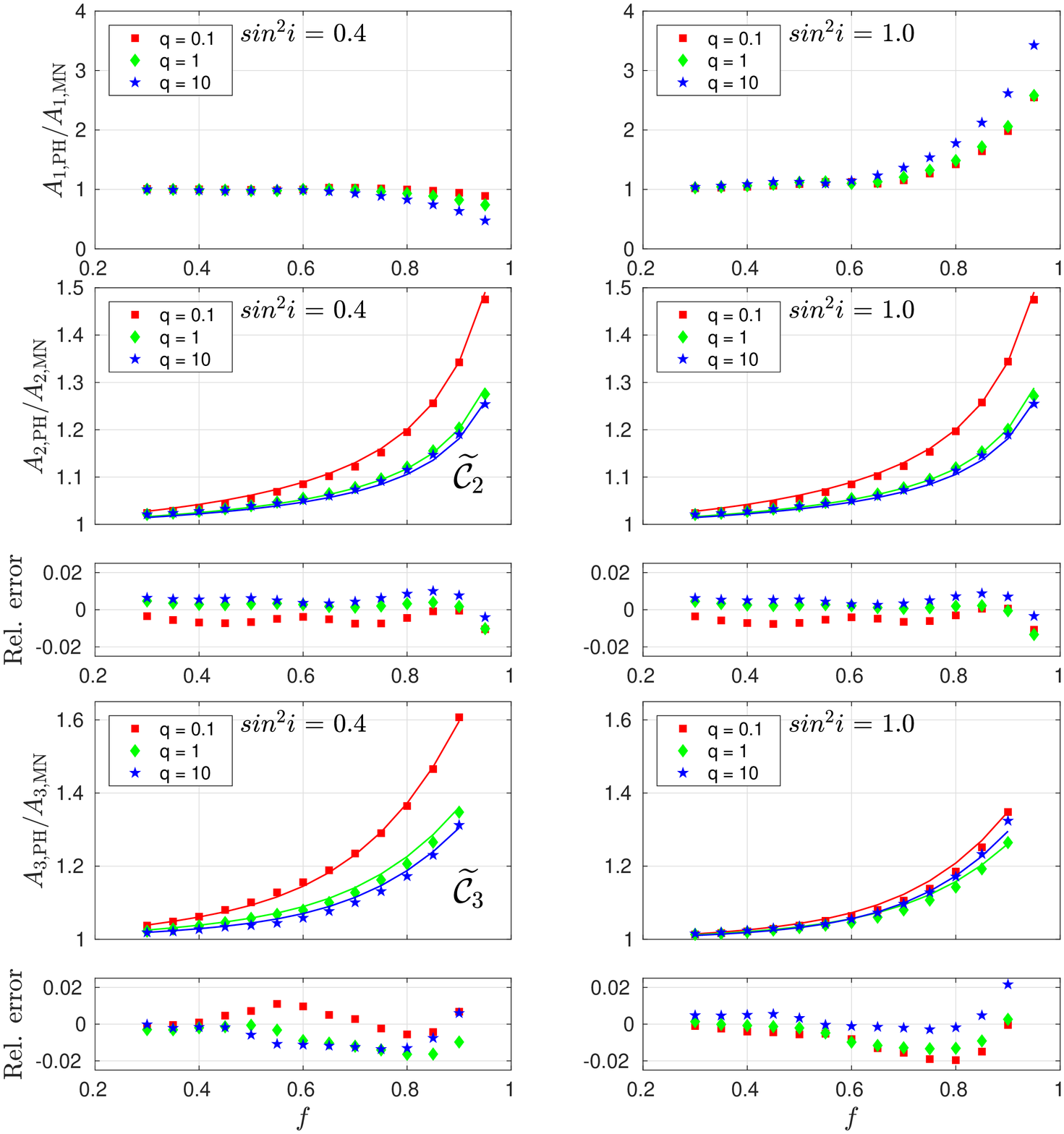}  }
	\caption{
		PHOEBE correction factors for the MN93 amplitudes.
		The two lower panels show second and third-harmonic correction factors (points) and their approximations (solid lines) as a function of fillout factor for  two inclinations and three mass-ratio values per inclination for a Sun-like star, with their {\it relative} residuals. Top panel shows first-harmonic correction factors for the same grid. Approximations (and residuals) are not available for the first harmonic; see text. 
	}
	\label{fig:CorrectionFactor}
\end{figure*}

In order to study the correction factor needed to be applied to the MN93 approximation, 
we calculated the correction factor on a linear 3D grid in $\log q$ ($-1.4$ to $2.0$ with step $0.2$), $f$ ($0.3$ to $0.95$ with step $0.05$) and $\sin ^2 i$ ($0.1$ to $1$ with step $0.1$).
For each point on the grid we  calculated the three amplitudes of the MN93 and PHOEBE models,
$\{A_{\rm n,\scaleto{MN}{5pt}};\, n = 1,2,3\}$ and $\{A_{\rm n,\scaleto{PH}{5pt}};\, n = 1,2,3\}$, 
and derived the correction factor 
%
\begin{equation}
C_n\left(q,f,i \right) = \frac{A_{\rm n,\scaleto{PH}{5pt}}}{A_{\rm n,\scaleto{MN}{5pt}}} \ 
\label{eq:corrfac}
\end{equation}
%
as a function of $q$, $f$ and $\sin i$.

To present the 3D correction factors we searched for a simple approximated expression for 
$\{C_n\left(q,f,i \right);\, n=1,2,3\}$,
using the EUREQA\footnote{https://www.nutonian.com/products/eureqa/} 
software package \citep{schmidt09,schmidt14}. Within EUREQA we searched for a separable expression of the form
\begin{equation}
\widetilde{C}_n \left(q,f,i \right) = 1 + F_{0}(q) F_{1}(f)  F_{2}(i) \ ,
\label{eq:sepcorr}
\end{equation}
that minimizes the relative error between the numerical ratio $C_n$ and $\widetilde{C}_n$. For the dominant second harmonic we found that
%
%
\begin{equation}
\widetilde{C}_2 \left(q,f \right) = 1 + \left( b_2 + \frac{c_2}{d_2+q}\right) \left(\frac{f}{a_2-f}\right) \ ,
\label{eq:anaformA2C}
\end{equation}
where
\begin{equation*}
a_2=1.0909 \ , \ b_2=0.0379 \ , \ c_2=0.0050 \ , \ d_2=0.0446 \ .
\label{eq:anaconstA2C}
\end{equation*}
%

The resulting approximation, with a maximum relative error of $2.4$\%, is independent of the inclination, which means that the ellipsoidal amplitude maintains its proportionality to $\sin ^2 i$ for systems with a large fillout factor.

For the third harmonic we found a  non-separable expression that resulted in a maximum relative error of $2.2$\% within $0.1 \leq q \leq 10$ and $f\leq0.9$,

\begin{equation}
\widetilde{C}_3 \left(q,f,i \right) = 1 +  
\frac{ (1+a_3 q\sin^2 i) f^6 + b_3 f^2 }{ (c_3+d_3\ln q) f + \sin^4 i }
\ ,
\label{eq:anaformA3C}
\end{equation}
where
\begin{equation*}
\ a_3=0.0698 \ ,\ b_3=0.2075 \ , \ c_3=2.0223 \ , \ d_3=0.3880 \ .
\label{eq:anaconstA3C}
\end{equation*}
%

Obviously, the two expressions yield a correction factor for every possible value of $f$, $q$ and $\sin i$. As expected, they converge to unity when the fillout factor approaches zero.

We could not find any simple expression for the correction of the first harmonic amplitude, so we give (see Appendix~\ref{app:A}) the corrected numerical values  as a linear interpolation between the calculated grid points through a Python code, available on GITHUB. 

The resulting amplitudes are presented in Fig.~\ref{fig:acPHOEBE} as a function of the fillout factor, for two inclinations and three mass-ratio values per inclination. 
The middle (lower) panel illustrates  $A_{\rm 2,\scaleto{PH}{5pt}}$ ($A_{\rm 3,\scaleto{PH}{5pt}}$) and its analytic approximation, given by $A_{\rm 2,\scaleto{MN}{5pt}} \cdot \widetilde{C}_2$ ($A_{\rm 3,\scaleto{MN}{5pt}} \cdot \widetilde{C}_3$). The behavior of the first harmonic coefficient $A_{\rm 1,\scaleto{PH}{5pt}}$ is shown in the upper panel of the figure.  

The amplitudes of all three harmonics rise monotonically with the fillout factor $f$. As can be seen in the figure, the second harmonic amplitude is much larger than the other two. 
For large enough values of $f$ and $\sin i$, its amplitude reaches a hundred ppts. The amplitude of the other two harmonics is an order of magnitude smaller.

Fig.~\ref{fig:acPHOEBE} shows that the analytical expression we derived for the second and third harmonic fits quite well the PHOEBE amplitudes. The derived correction factors themselves are shown in Fig.~\ref{fig:CorrectionFactor}. 
The  middle (lower) panel shows the correction factor $C_2$  ($C_3$)  and its approximations 
with the same grid used before. 
The correction factor starts at  $1$ for $f = 0$ (no correction), as expected, and rises monotonically as $f\to 1$, obtaining a maximum value of $\sim$ 1.5 at $f \gtrsim 0.90$.

The residual part of each panel shows the {\it relative} residuals of the correction factor according to the derived expression compared to the measured one. As stated, the residual panels show a typical relative error $<1\%$. The maximum error, of $\sim$$2\%$, is
obtained for the case of $C_2$, when the fillout factor is large.

The top panel of Fig.~\ref{fig:CorrectionFactor} presents the correction factor $C_1$, which  behaves differently than $C_2$ and $C_3$. For edge-on binary,  $C_1$ rises monotonically up to $\sim$ 3 at $f = 0.95$, but {\it decreases} as a function of $f$ for $\sin i=0.4$, down to a value of $0.5$.  

%


%

\subsection{Extension of the  approximated correction}
So far, the correction factor ${C}_1$ and the approximated expressions, $\widetilde{C}_2$ and $\widetilde{C}_3$, were calculated for 
V-band light curves of a Sun-like primary system. To explore our approach on systems with different primaries and different observational bands we tested 
them for main-sequence stars of $0.8$, $1$, $2$ and $5$ $M_{\odot}$, over the Johnson B, V, and R, and Cousins-I bands, with different $f$, $q$ and $\sin i$ values. 

We found that the correction factor of the first harmonic ${C}_1$ changes dramatically for binary systems of different primaries and different observational bands. Therefore, the first-harmonic amplitudes were calculated numerically over a grid of $f$, $q$ and $\sin i$, for different main-sequence stars and optical bands, using PHOEBE. These can be extracted by the  Python module described in Appendix~\ref{app:A}. The module derives the estimated amplitudes for any binary parameters by interpolating between the grid points.

Even better, in all our tests the approximated  expressions $\widetilde{C}_2$ and $\widetilde{C}_3$ fitted well, by up to
$\sim 7$\% ($\sim17$\%) for $\widetilde{C}_2$ ($\widetilde{C}_3$).
These results show that our approximation for the second and third harmonic is reliable and robust, and can be used for different stars and different bands. 

\subsection{Limits on the use of the correction factors}

The above analysis and the resulting correction factors were based on a few  assumptions, such as a circular, aligned and synchronous orbit, and our knowledge of the  limb and gravity-darkening coefficients. In addition, the focus of the analysis on the ellipsoidal effect ignores the reflection of the primary light by the secondary surface, and stellar spot modulations, for example. Here we try to estimate and quantify the limits on physical parameters within which this analysis is still valid.

\begin{itemize}
\item An eccentric orbit: From Equation~(\ref{eq:ellip1}) we see that for a small-eccentricity orbit the ellipsoidal leading amplitude $A_2$ will deviate from its circular-orbit value by a factor of $\sim$$(1+e)^3 \simeq 1+3e$, where $e$ is the eccentricity \citep{engel20}. This means that our analysis and the resulting corrections factors are valid as long as $3e \ll 1$.

\item Limb and gravity-darkening coefficients: Our analysis is based on estimation of the limb and gravity-darkening coefficients, both expected to be in the 0--1 range.
 From the $\alpha_2$ formula in Equation~(\ref{eq:alpha}) we see that $A_2$  is mainly sensitive to the gravity-darkening parameter $\tau$, so that $A_2 \propto (1+\tau)$. 
  Due to uncertainties in $T_{\rm eff}$, $\log g$ and stellar models, we expect the uncertainty in the gravity-darkening  coefficient, $\Delta \tau$, to be ${\Delta \tau}/{\tau} \lesssim 0.1$.
 Thus, we suggest that our analysis is almost not effected by $\Delta \tau$, since 
${\Delta A_2}/{A_2} \simeq {\Delta \tau}/{(1+\tau)} \ll 1$.
\end{itemize}

The reflection contribution can be estimated as
 $\sim$$- p_{\rm geo}\left(\frac{R_2}{a}\right)^2 \sin i \cos\phi$, where $R_2$ is the secondary radius and $p_{\rm geo}$ is the geometric albedo, expected to be in the 0--0.5 range \citep[e.g.,][]{faigler11}. Thus the reflection modulation is negligible relative to the first harmonic amplitude $A_1$, if $p_{\rm geo} \left(\frac{R_2}{a}\right)^2 \sin i \ll \left|A_1 \right|$.


\section{One example: OGLE-BLG-ELL-007730}
\label{sec:test}

\subsection{The observed ellipsoidal modulation}

\begin{figure*} 
 	\centering
 	\resizebox{18cm}{8cm}
 	{ 		\includegraphics{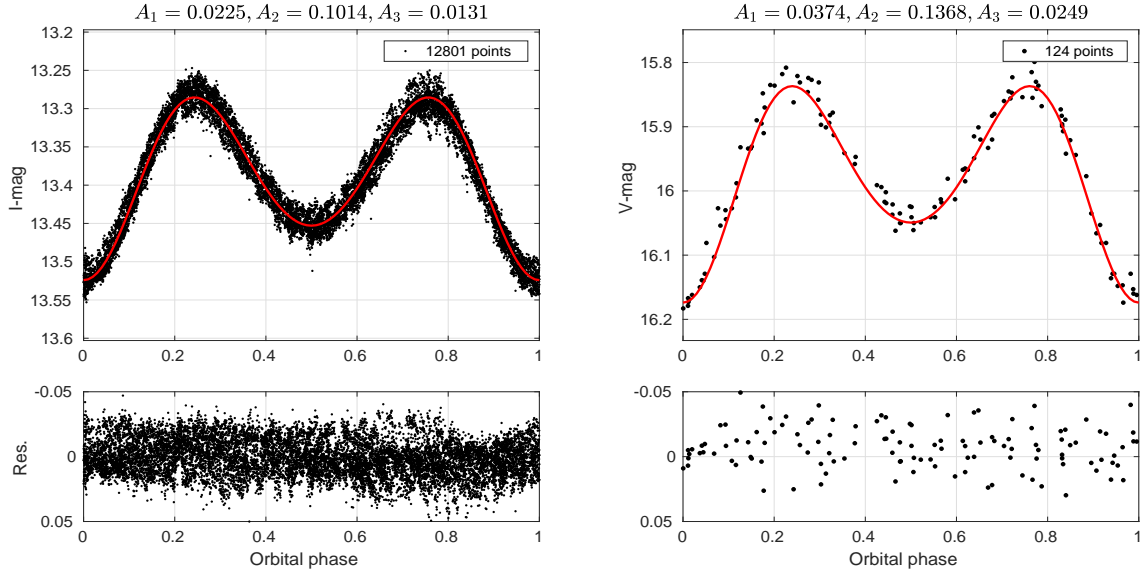}   	}
 	\caption{
Folded OGLE-IV light curve of OGLE-BLG-ELL-007730 in the $I$ (left panel) and $V$ (right panel) bands. The orbital phase is calculated with a period of $24.558264$ d and  zero phase at HJD $2457019.8283$. 
A three harmonics model is plotted with a solid line and the fitted coefficients are given in the top part of the figure. 
The predominant $A_2$ coefficient gives the characteristic double-peaked appearance to the light curve, while $A_1$ and $A_3$ contribute to the difference between the minima. The residuals are plotted in the lower panels.
}
\label{fig:ogle}
 \end{figure*}

As an example of an observed ellipsoidal modulation, we show in Fig.~\ref{fig:ogle} the $V$ and $I$ OGLE-IV light curves of OGLE-BLG-ELL-007730 \citep{soszy16}, 
folded with the OGLE derived orbital period P = $24.558264$ d and $\phi=0$ at a $T_0$ = HJD $2457019.8283$. The system was chosen from the public OGLE Collection of Variable Stars\footnote{http://ogledb.astrouw.edu.pl/$\sim$ogle/OCVS/} by searching for a system with a large relative ellipsoidal modulation and a significant difference between the two minima, as seen in the figure.

Overplotted in the figures are three-harmonic models with
amplitudes of $A_1 = 0.0374 \pm 0.0025$, $A_2 = 0.1368 \pm 0.0025$ and $A_3 = 0.0249 \pm 0.0024$ for $V$ and $A_1 = 0.02251 \pm 0.00018$, $A_2 = 0.10140 \pm 0.00018$, $A_3 = 0.01310 \pm 0.00018$ for the $I$ band. The residuals are plotted in the lower panels.

%

\subsection{Estimating the mass ratio}
Equipped with our corrected estimation for the ellipsoidal modulation, we now use OGLE-BLG-ELL-007730 as an example of how our search for dormant BH might work. 

\begin{figure*} 
	\centering
	\resizebox{12cm}{9cm}
	{		\includegraphics{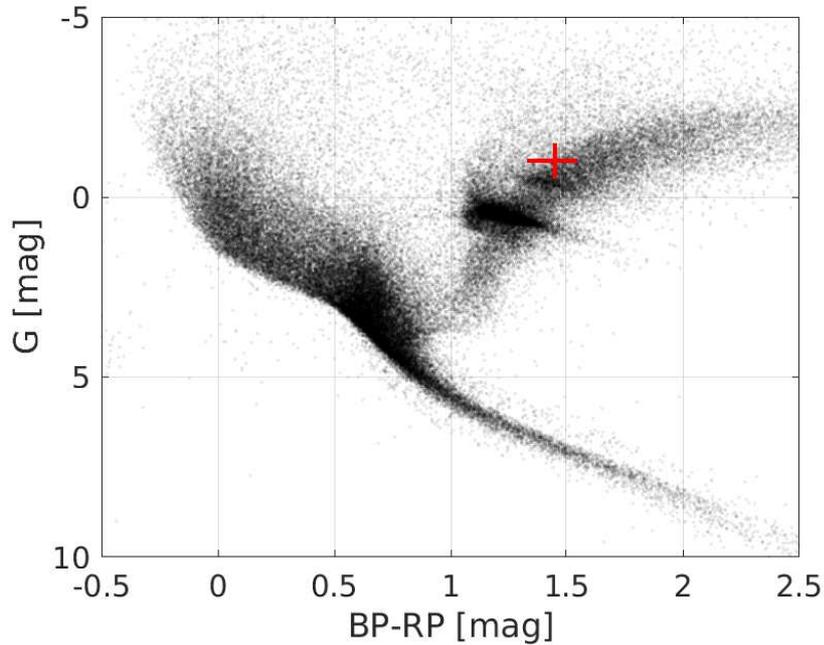} 	}
	\caption{ OGLE-BLG-ELL-007730 on the Gaia CMD; see text. As a background, we plotted a grey-scale density map of Hipparcos stars, used as a proxy for the
expected CMD in the solar neighbourhood, as done, for example, by \citet{SM19}.
}
	\label{fig:CMD}
\end{figure*}

According to Gaia DR2 \citep{GaiaDR2}, the stellar parallax is $0.213 \pm 0.052$, its G magnitude is $14.7300 \pm 0.0066$, with an extinction of 
$A_G \simeq 2.4\pm 0.2$, BP-RP $=2.641 \pm 0.033$ and reddening of E(BP-RP) $\simeq 1.19\pm 0.12$. We use all these values to locate the star on the Color-Magnitude Diagram (CMD) in Fig.~\ref{fig:CMD}, indicating that OGLE-BLG-ELL-007730 is either on the ascending giant branch or maybe an asymptotic giant branch star. 

The TIC\footnote{https://tess.mit.edu/science/tess-input-catalogue/} 
\citep{TIC}
stellar values are $T_{\rm eff}$ of $4200$K and stellar radius of $26\, R_{\odot}$, consistent with the CMD position. If we assume a typical giant mass of $\sim1.5\, M_{\odot}$, we obtain $\log g\sim1.8$. 

\begin{figure*} 
	\centering
	\resizebox{18cm}{9cm}
	{		\includegraphics{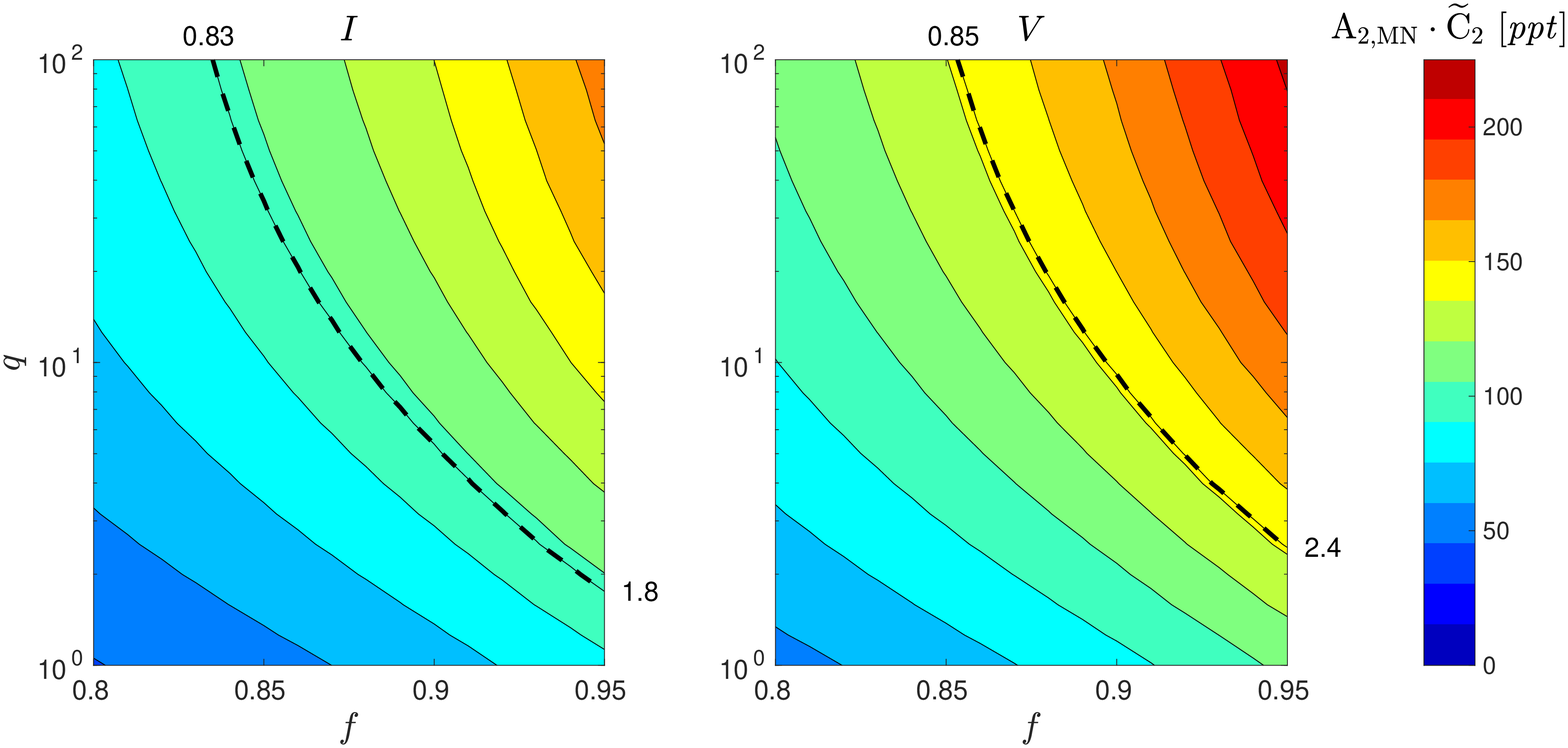} 	}
	\caption{ 
		$I$ (left) and $V$ (right) amplitudes of the second harmonic of the ellipsoidal effect as a function of $q$ and $f$, based on the derived corrections (see above).
		Contour lines with equally-spaced amplitudes are drawn with solid lines from $15$ ppt upwards in steps of $15$ ppt. 
		The locus of OGLE-BLG-ELL-007730 is plotted in both panels, and its crossing points with both axes are given.
	}
	\label{fig:ogle_qf}
\end{figure*}

With these stellar details, we can now plot the corrected  amplitudes of the second harmonic expected for the ellipsoidal modulations in the $V$ and the $I$ bands of OGLE-BLG-ELL-007730.  This is done in Fig.~\ref{fig:ogle_qf} as a function of the Roche-lobe fillout factor of the primary $f$ and the binary mass-ratio $q$ . The figure was derived for $\sin i=1$, with no contribution from the secondary and no beaming effect, using the linear limb- and gravity-darkening coefficients of \cite{claret11},
$u$ = $0.638$ ($0.845$) and $\tau$ = $0.400$ ($0.567$) for the $I$ ($V$) band, and assuming zero metallicity. 

%
%

The combination of the two loci suggests a range of mass ratios of $2.4\leq q\leq 100$, indicating a binary with a mass ratio larger than unity. This conclusion is based only on the light curve modulation and the estimation of the stellar temperature. 
We propose that the actual mass ratio of OGLE-BLG-ELL-007730 is close to the small end of this range, $2.5 \lesssim q$, not untypical of an Algol binary, for which the less massive star is the evolved one, like here \citep[e.g.,][]{nelson01}.  For such a mass-ratio value, Fig.~\ref{fig:ogle_qf} suggests that  $f\sim0.9$, indicating that the evolved star is close to filling its Roche lobe. In such a case, the amplitudes of the third harmonic are expected to be $\sim20$ and $\sim10$ ppts for the $V$ and $I$ bands, similar to the values derived from the OGLE light curves.
Obviously, solving the system requires a detailed analysis of the light curves in both bands, given the stellar radius and its mass.

\section{Summary and Discussion}
\label{sec:summary}

This study presents a correction to the MN93 analytical approximation of the ellipsoidal modulation, based on the PHOEBE numerical code. We derived corrections for the amplitudes of the first three harmonics of the modulation  for a 3D grid, as a function of the mass ratio, the inclination of the binary and 
the fillout factor of the primary---the ratio between the stellar radius and the radius of its Roche lobe. The correction can get up to a factor of $\sim1.5$ when the star is close to filling its Roche lobe. We present simple expressions for the correction of the second and third harmonics. A Python code to extract the three-harmonics coefficients is given. 

The combination of the MN93 approximation and our correction can be used to obtain the expected ellipsoidal modulation for a close binary system, provided some simplified assumptions of the model are fulfilled. These include that the star is stationary in the rotating frame, with no differential rotation, and that the stellar atmospheric models apply to distorted stars, to mention two examples.

Note that we consider here only the ellipsoidal modulation of the primary induced by the secondary, and ignore the other well-known two effects: reflection/emission and beaming. The reflection (emission) modulation is the result of light coming from one component and reflected (absorbed and thermally emitted) by the other one
\citep[e.g.,][]{vaz85, maxted02, for10, faigler11}. The relativistic beaming effect causes the intensity of a light source to increase (decrease) when the source is moving towards (away from) the observer
\citep[e.g.,][]{rybicki79, loeb03, zucker07, mazeh10, bloemen11, eigmuller18}. Obviously, any complete analysis of a light curve has to account for the other two modulations too. 

Fortunately, whereas most of the variability of the ellipsoidal modulation appears in the cosine function of the second harmonic, the beaming modulation for a circular orbit appears in the sine function of the first  harmonic,
and most of the variability of the reflection effect is concentrated at the cosine function of the first harmonic. Therefore, in principle, the reflection and beaming effects can be separated from the ellipsoidal modulation in the analysis of the light curves.  

The separation between the ellipsoidal effect on one hand and the beaming and reflection modulations, on the other hand, is not possible for a binary with an eccentric orbit. 
First, the shape and amplitude of the ellipsoidal effect depend on the eccentricity and argument of periastron of the orbit, and its power is not necessarily concentrated in the second harmonic. Second, the beaming and reflection effects  have power in all three first harmonics, with the sine and cosine functions alike \citep[see][for a detailed model, eBEER, for eccentric orbit]{engel20}.
Therefore, the corrections presented here can be applied for circular or nearly circular orbits only. Furthermore, 
our analysis assumes that the binary is in a tidal equilibrium state, so the system has reached not only circularization but also synchronization and alignment of the stellar rotation with the binary orbital angular momentum. 

However, as shown by many studies 
\citep[e.g.,][]{mayor84, mathieu88, mazeh08},
most of the short-period binaries have reached tidal equilibrium, and the orbits have small eccentricities only.
 In fact, tidal equilibrium is probably reached for any binary with a large fillout factor of the primary. 
This is relevant here, as this search for dormant BHs targets short-period binaries with a primary with a large fillout factor, because only such systems display large enough ellipsoidal modulation to be detected and identified as such.  For those systems the corrections we developed here are relevant.  
 Furthermore, our analysis is applicable also for binaries that have reached tidal equilibrium.

 We note in passing that we are participating in an effort to extend the BEER (BEaming Ellipsoidal and Reflection) approximation for which the ellipsoidal model uses MN93 to eccentric orbits. A paper (Engel et al.) summarizing this work was submitted to MNRAS. The next stage, out of the scope of the current paper, is to combine the present analysis with the Engel et al. approach to construct a modified model for eccentric binaries, even for large fillout factor systems.

Using the corrected amplitudes of the harmonics of the ellipsoidal modulation of a short-period binary can help to reach a reliable estimate of the binary mass ratio. At the first stage, one obtains a constrain on the binary mass ratio, fillout factor and inclination. In some cases, like OGLE-BLG-ELL-007730, one can obtain a minimum mass ratio that is larger than unity. Provided the mass and radius of the primary and the binary period are known, one can solve for the three unknowns. 

Obviously, when analyzing a specific light curve one has to estimate the contribution of the secondary star to the brightness of the system.
A good candidate for having a compact secondary is a system that the analysis suggests the secondary is faint {\it and} is more massive than the primary. 
Then, of course, one has to show that the secondary is not a main-sequence star. This is especially true for systems with giant or sub-giant primaries, as is OGLE-BLG-ELL-007730, for which the obvious conjecture is that the system is an Algol-type binary. In such systems, which are relatively frequent \citep[e.g.,][]{budding04}, the main-sequence star is indeed more massive that the optical primary. 

The analysis  proposed here can be instrumental in searching, identifying, and analyzing ellipsoidal variables in large data sets which are available already, and the ones coming us soon, like the light curves collected by LSST, which will start  operating in the near future \citep{abell09,ivezic19}.
As will be demonstrated in the future papers of this series, to confirm the binarity of a system and its large mass ratio, such a system has to be followed by radial-velocity (RV) observations. Existing and coming up multi-object spectrographs: VIMOS \citep{leFevre03}; FMOS \citep{maihara00}; GIRMOS \citep{wright00}; OSMOS \citep{stoll10}; GMACS \citep{dePoy12}; DEIMOS \citep{faber03}; LAMOST \citep{su98}; 4MOST \citep{deJong11};
can be used to follow up many candidates, as was done, for example, by,  \citet{tal-or15}, \citet{romani15} and \citet{rebassa17}.

The photometric modulation is not the only approach to discover dormant BHs in binaries.
Single-lined spectroscopic binaries present RV variations, induced by their unseen companions. Given the orbital parameters, one can derive the binary mass function, and obtain a minimum of the secondary mass, provided the primary mass can be estimated. This is a simple way to identify massive companions, some of which can be dormant BHs.

The ellipsoidal technique is sensitive to systems with a primary that fills most of its Roche-lobe, and therefore is limited to short-period binaries, depending on the stellar radius. The RV technique, on the other hand, can be applied to longer-period binaries. Indeed, two studies claimed  recently finding dormant BHs in such spectroscopic binaries. \citet{liu19} announced the discovery of a B-type primary star with a BH companion of $68\, M_{\odot}$, moving in a relatively long orbital period of $\sim79$ days. This system has been found in the RV monitoring campaign of LAMOST \citep{cui12} to discover and study spectroscopic binaries. Another work, by \cite{thompson19}, reported the discovery of a BH candidate of $3.3\, M_{\odot}$ 
and a giant-star binary system, with an orbital period of $\sim83$ days, found while searching for binary systems with massive unseen companions in the  APOGEE spectroscopic data  \citep{majewski17}.

However, there are still some doubts about at least one of these detections. 
\cite{elbadry20} showed that the H$\alpha$ line in LB-1, the binary reported to contain a BH companion of
 $68\, M_{\odot}$, has non-significant RV variability. This undermines the derived mass ratio of the system and thus the reported unprecedentedly high-mass companion. Instead, a normal-mass BH seems more plausible. \cite{eldridge19} and \cite{irrgang20} proposed that the luminous star could be a $\sim$ $1\, M_{\odot}$ pre-subdwarf in a short-lived evolutionary phase. In this scenario, the companion would have a significantly lower mass and could even be a neutron star.

One should note that the BH binaries are probably quite rare. As long as we examine relatively small samples, of the order of $\sim10^5$ stars, like the LAMOST or APOGEE ones, the chances of finding in their binary samples \citep[e.g.,][]{yi19,price-whelan20} a real BH binary is small. On the other hand, the ellipsoidal technique can survey very large samples, of the order of $\sim10^8$ stars with precise enough light curves, and therefore might have a better chance of finding BH binaries. This depends, of course, on how the frequency of the BH binaries varies as a function of binary period, an unknown statistical feature that depends on the evolutionary tracks of the binaries that led to the formation of the BH binaries \citep[e.g.,][]{Wiktorwicz19}.  

One obvious major step to understand the population of the BH binaries will occur when Gaia will release their astrometric measurements. These data will allow identifying BH binaries with a preference for binaries with longer periods \citep[e.g.,][]{andrews19}, on the order of a few years, with an approach outlines by \citet{shahaf19} and \citet{belokurov20}, for example. The combination of the three techniques, using ellipsoidal effect, RV modulation and astrometric motion, will finally give us the information needed to  obtain  the BH binaries frequency and its dependence on the binary period. 

\newpage

\section*{Acknowledgments}
%

We are indebted to the referee, Andrej Pr{\v{s}}a, who contributed illuminating comments and suggestions on the previous version of the manuscript, helping us improve substantially the paper. We acknowledge the usage of PHOEBE \citep{prsa05,prsa16,horvat18} and deeply thank its support team, especially Andrej Pr{\v{s}}a and Kyle Conroy, for their endless effort to help us run the software. We could not have completed this work without their prompt and kind support. We thank Michael Engel for his kind assistance in building our binary model in PHOEBE
and Sahar Shahaf, Laurent Eyer, Pierre Maxted, Igor Soszy\'nski and Micha{\l} Pawlak for valuable remarks.
We acknowledge the usage of EUREQA \citep{schmidt09,schmidt14}.
This research was supported by Grant No.~2016069 of the United States-Israel Binational Science Foundation (BSF) and by the Israeli Centers for Research Excellence (I-CORE, grant No.~1829/12).

\section*{Data availability}

The data underlying this article are available in the article and in its online supplementary material.


\bibliographystyle{mnras}
\bibliography{Aellip_bib}

%

\newpage

\appendix
\section{Phython module to derived the corrected ellipsoidal amplitudes}
\label{app:A}

We composed an online Python module\footnote{https://github.com/roygomel/EllipsoidalAmplitudes} that calculates the expected semi amplitudes of the first three harmonics, based on the PHOEBE simulated light curves. The code interpolates the amplitudes values for the first harmonic using the grid we prepared and uses Equations~\ref{eq:anaformA2C}--\ref{eq:anaformA3C} for the second and third harmonics.

The code inputs are the estimated effective temperature, 
log gravity, metallicity and the Roche-lobe fillout factor of the primary star, the inclination and mass ratio of the binary and the observing band.
The code range is given in Table~\ref{tab:Ai}.

\renewcommand{\arraystretch}{2}
\begin{table*}

	\large \begin{tabular}{|c|c|c|c|}
		\hline
		 & $A_{\rm 1}$ &  $A_{\rm 2}$ &  $A_{\rm 3}$ \\
				
		\hline
		fillout factor  		& f $\leq0.9$ & f $\leq0.95$ & f $\leq0.9$ \\
		Mass ratio  			& $0.05\leq$ q $\leq10$ & $0.05\leq$ q $\leq100$ & $0.1\leq$ q $\leq10$ \\
		Orbital inclination  	& $0.2\leq \sin^2 i \leq1$ & No limitation & No limitation \\
		Primary mass [$M_{\odot}$] 	& $0.8\leq M_1 \leq5$ & No limitation & No limitation \\
		Evolutionary state  	& Main sequence & No limitation & No limitation \\
		Band  					& $B$, $V$, $R$, $I$ & * & * \\		
		\hline 
		 
	\end{tabular} 
	\\
	\vspace{3mm}
	$^*$All available bands in \cite{claret11}.
	\caption{The Python module range of operation.}
	\label{tab:Ai}
\end{table*} 

\end{document}